\documentclass[12pt,aaspp4,psfig]{article}
\usepackage{aaspp4}

\def\N#1|#2{N\ifx.#2\else_{\strut{\rm\ifnum#2=0\relax null\ true\else null\ false\fi}}\fi\ifx.#1\else^{\strut{\rm\ifnum#1=0\relax maintain\else reject\fi}}\fi}
\newcommand\FDR{{\rm FDR}}

\begin{document}

\title{Controlling the False Discovery Rate\\in Astrophysical Data Analysis}

\author{Christopher J. Miller} 
\affil{Dept. of Physics, Carnegie Mellon University, 5000 Forbes Ave., Pittsburgh, PA-15213}

\author{Christopher Genovese}
\affil{Dept. of Statistics, Carnegie Mellon University, 5000 Forbes Ave., Pittsburgh, PA-15213}

\author{Robert C. Nichol} 
\affil{Dept. of Physics, Carnegie Mellon University, 5000 Forbes Ave., Pittsburgh, PA-15213}

\author{Larry Wasserman}
\affil{Dept. of Statistics, Carnegie Mellon University, 5000 Forbes Ave., Pittsburgh, PA-15213}

\author{Andrew Connolly}
\affil{Dept. of Physics \& Astronomy, University of Pittsburgh, 3941 O'Hara Street, Pittsburgh, PA-15260}

\author{Daniel Reichart}
\affil{Dept. of Astronomy, CalTech, 1201 East California Blvd, Pasadena, CA-91125} 

\author{Andrew Hopkins}
\affil{Dept. of Physics \& Astronomy, University of Pittsburgh, 3941 O'Hara Street, Pittsburgh, PA-15260}

\author{Jeff Schneider, Andrew Moore}
\affil{School of Computer Science, Carnegie Mellon University, 5000 Forbes Ave., Pittsburgh, PA-15213}

\begin{abstract}
The False Discovery Rate (FDR) is a new statistical procedure to 
control the number of mistakes made when performing multiple hypothesis
tests, {\it i.e.} when comparing many data against a given model hypothesis.
The key advantage of FDR is that it allows one to {\it a
priori} control the average fraction of false rejections made (when comparing to the
null hypothesis) over the total number of rejections performed. We compare FDR
to the standard procedure of rejecting all tests that do not match the null hypothesis above
some arbitrarily chosen confidence limit, {\it e.g.}
$2\sigma$, or at the 95\% confidence
level. We find a similar rate of correct detections,
but with signicantly
fewer false detections. Moreover,
the FDR procedure is quick and easy to compute and can be trivially adapted
to work with correlated data. The purpose of this paper is to introduce the FDR
procedure to
the astrophysics community.
We illustrate the power of FDR through several astronomical examples,
including the detection of features against a smooth one-dimensional function,
{\it e.g.} seeing the ``baryon wiggles'' in a power spectrum of matter fluctuations,
and source pixel detection in imaging data.
In this era of large datasets and high precision measurements,
FDR provides the means to adaptively control a scientifically meaningful quantity --
the number of false discoveries made conducting multiple hypothesis tests.
\end{abstract}


\section{Introduction}

A recurrent statistical problem in astrophysical data analysis is to decide
whether data are consistent with the predictions of a theoretical model.
Consider, as an example, the comparison between an observed power spectrum (e.g., of galaxies or clusters),
where each data point gives a noisy estimate of the true power spectrum at a single wave number,
and the functional form hypothesized by a specific cosmological model.
One can test for overall differences between data and model
using a statistical measure of discrepancy,
such as a simple $\chi^2$ if the data are uncorrelated.
If this discrepancy is sufficiently large, 
we conclude that there are ``significant'' differences beyond those accounted for by randomness in the data.
This is an example of a statistical hypothesis test.

However, this test indicates only whether the data and model differ overall;
it does not specify where or how they differ.
To address such questions, a single test is not enough.
Instead, one would need to perform multiple hypothesis tests,
one at each wave number, based on the discrepancy between the data 
and model spectrum at that wave number.
It is common, for
example, to declare a test significant if the discrepancy is greater than twice
the standard error of the measurement; we call this the ``$2\sigma$''
approach.
This rule is calibrated to declare significance erroneously with probability about 0.05.
However,
the probability of making such errors increases rapidly with the number of tests performed,
so this calibration will typically be inappropriate for multiple testing.
One can adjust, in part, by making the tests more stringent at each wave number,
such as using a ``$3\sigma$'' cut off.
Unfortunately,
while this does reduce the probability of spurious detections,
it also reduces (often severely) the probability of correctly detecting real deviations,
especially those on the edge of detectability that can be most interesting.

The need to perform multiple hypothesis tests, and the attendant difficulties,
are ubiquitous in astronomy and astrophysics.
In this paper, we present an effective method for multiple testing
that improves the probability of correct detections over methods in current use
while still controlling the probability of spurious detections.
The method, 
due to Benjamini \& Hochberg (1995) and the subject of much recent research in the statistical literature,
bounds a particular measure of inaccuracy called the False Discovery Rate (FDR).

We stress here that FDR is not a new testing technique
but rather a method for combining the results of many tests of any kind.
FDR is still relatively new in the statistical literature, but it has
already been shown to possess several key advantages over existing methods:
\begin{itemize}
\item It has a higher 
probability of correctly detecting real deviations between model and data.
\item It controls a scientifically relevant quantity -- the average fraction
of false discoveries over the total number of discoveries.
\item Only a trivial adjustment to the basic method is required to handle correlated data.
\end{itemize}

In Section \ref{sec::mult-test}, we review statistical hypothesis testing
and discuss the problems that arise when multiple tests are performed.
We illustrate these ideas with the example of image source detection:
how does one decide which pixels in a CCD image belong to the sky background
and which are part of a source?
In Section \ref{sec::fdr}, we describe the FDR method in detail.  
Section \ref{sec::sims} gives simulation results based on the
source-detection example that compare and contrast the available methods. 
Section \ref{sec::phys-examp} describes how FDR is applied in a variety of other astrophysical examples.
Finally, Section \ref{sec::disc} discusses the role of FDR in astrophysical data analysis.
We present a heuristic proof of the FDR procedure in Appendix \ref{sec::fdrproof} and
a step by step worked example in Appendix \ref{sec::wexamp}.

\section{Multiple Hypothesis Testing}\label{sec::mult-test}

Consider a common astronomical problem: source detection in images.  
Each pixel in the image can be thought of as being mostly part of the background or
mostly part of a source.
We will call these ``background'' and ``source'' pixels, respectively.
To illustrate hypothesis testing,
we will focus in this section on the simplified problem of
deciding, for each pixel, whether it is background or source.
We will take up the full details of source detection 
(\emph{e.g.}, identifying stars and galaxies from the source pixels)
in a future paper (Hopkins et al., in preparation).

The data in the source detection problem is a measured photon count at every pixel.
We expect the counts for source pixels to be larger (on average) than the counts for
background pixels.
Hence, we select a critical threshold and classify a pixel as source (background)
if its count lies above (below) that threshold.
Such a rule for classifying a pixel using the data is an example of a \emph{statistical hypothesis test}.
We are deciding between two competing hypotheses: the \emph{null hypothesis} that
the pixel is background and the \emph{alternative hypothesis} that the pixel is source.\footnote{Technically,
there is a family of alternative hypotheses indexed, for example, by the true source intensity,
but this does not change how we construct the test.}
The test is performed by comparing a \emph{test statistic} to a \emph{critical threshold}.
The test statistic is a function of the data whose probability distributions
under the null and alternative hypotheses are as separated as possible.
In source detection,
the test statistic might be, for example, the photon count minus the estimated
background, divided by the standard deviation of the background counts.
The choice of critical threshold determines the statistical properties of the test,
as we will discuss below.
If the test statistic is above the threshold, we \emph{reject} the null
hypothesis in favor of the alternative.  Otherwise, we \emph{maintain} the
null hypothesis because it is either true or because we cannot detect that it
is false with the available data.
\footnote{One sometimes hears the phrase ``accept the null hypothesis'', but this is inappropriate,
and possibly dangerous.
We can fail to reject the null hypothesis for two reasons:
because it is true or because the data are insufficient or too noisy
to detect the discrepancy.
It is typically not possible to distinguish between these two cases.
The term ``accept'' implies that the data support the null, which they need not.
The null hypothesis acts as a working model, a status quo;
we reject the null hypothesis when its predictions imply that the given data are
exceedingly unlikely to have been observed.
For this reason, we prefer the phrase ``maintain the null hypothesis''.}

\begin{figure}[p]
\epsfbox{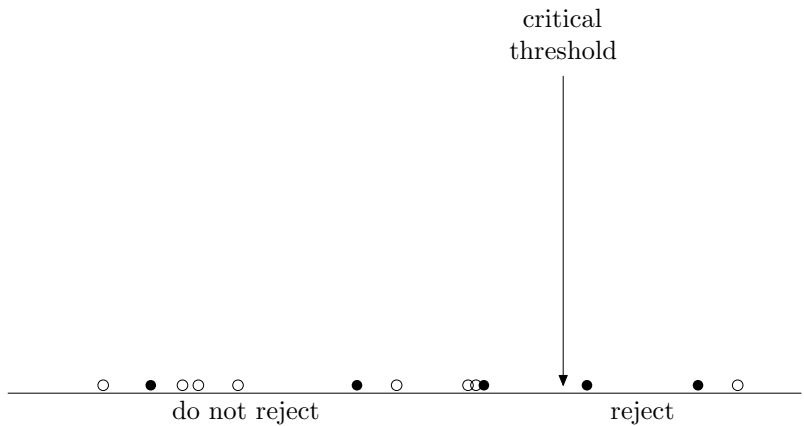}
\caption[]{\label{fig::TypeError} Schematic illustration of the two types of errors that can be made in hypothesis testing. 
The circles correspond to the values of a test statistic for several hypothesis tests;
they are filled or not according to whether the null hypothesis is false or true for that test.
The test rejects the null hypothesis if the value of the test statistic lies to the right of the critical threshold.
Filled circles to the left of the threshold are false non-rejections (Type II errors).
Unfilled circles to the right of the threshold are false rejections
(Type I errors, false discoveries).}
\end{figure}

In practice, no matter what the threshold, we will incorrectly classify some pixels
(e.g. false positives or false negatives).
We graphically represent this in Figure \ref{fig::TypeError}, where a threshold is set
and null hypotheses with test statistics to the right of this threshold are rejected. In this example,
we have mistakenly rejected one true null hypothesis, while we have maintained (to the left
of the threshold)
three real sources (null hypothesis false). This illustrates
the two types of errors we can make:
(i)~incorrectly identifying a background pixel as source
and 
(ii)~incorrectly identifying a source pixel as background.
In the first case, we have erroneously \emph{rejected} the null hypothesis when it is true.
In the second case, we have erroneously maintained the null hypothesis when it is false.
Statisticians refer to these (memorably) as Type I and Type II errors respectively.
Generally, statisticians talk about \emph{power} rather than Type II error;
the power of a test is the probability of rejecting the null hypothesis given that it is false.
The power is thus one minus the probability of a Type II error.
One wants the power to be as high as possible,
but there is a trade-off between power and the probability of a Type I error:
reducing one raises the other.

\begin{figure}[p]
\epsfbox{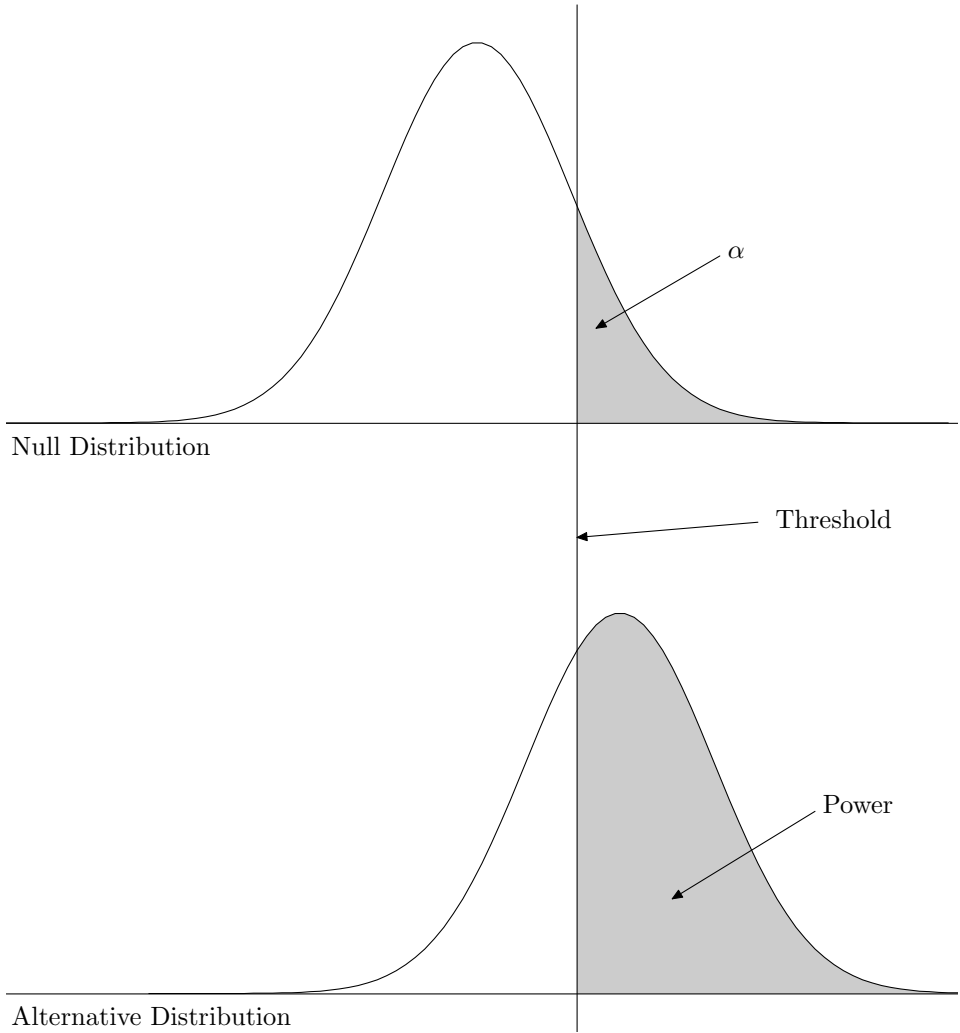}
\caption[]{\label{fig::power} Schematic illustration of the trade-off 
between the probability of Type I error and the power.
The top curve is the distribution of the test statistic when the null hypothesis is true;
the bottom curve is the distribution of the test statistic when one alternative
hypothesis is true.
The vertical line through both curves represents
the selected critical threshold for the test.
The area under each curve to the right of the critical threshold represents the probability
of rejecting the null hypothesis in the corresponding case.
When the null hypothesis is true, it is the probability of making a Type I error, $\alpha$.
When an alternative hypothesis is true, it is the power under that alternative.}
\end{figure}

The trade-off between minimizing Type I error and power is
illustrated in Figure \ref{fig::power}. In the top panel, we 
show the null  distribution (i.e. the real background).
We reject anything to the right of the threshold (vertical line) as a source.
We will mistakenly identify the background data above the threshold as real sources, since our test rejects
those data. 
As we raise the threshold (for a fixed null distribution), we minimize the 
number of Type I errors (false discoveries). In the bottom panel, we show the source (alternative)
distribution. We identify sources as those data which lie to the right
of the same threshold (meaning that we will fail to identify real sources to the left of
the threshold).  If we raise the threshold (which reduces false discoveries), we
identify fewer real sources, i.e. there is a trade-off between the number of
false discoveries and the power.

Throughout this paper, we use the terms ``rejection'', ``discovery'',
``detection'', and ``event'' interchangeably,
based on what is clearest in context.
Thus, 
a Type I error can be described as a ``false discovery'' or a ``false detection''.
Similarly, we focus on power rather than the complementary probability of Type II errors,
so we write ``correct discovery'' or ``correct detection''
for cases where the null hypothesis was rejected when it is false.

\begin{figure}[p]
\epsfbox{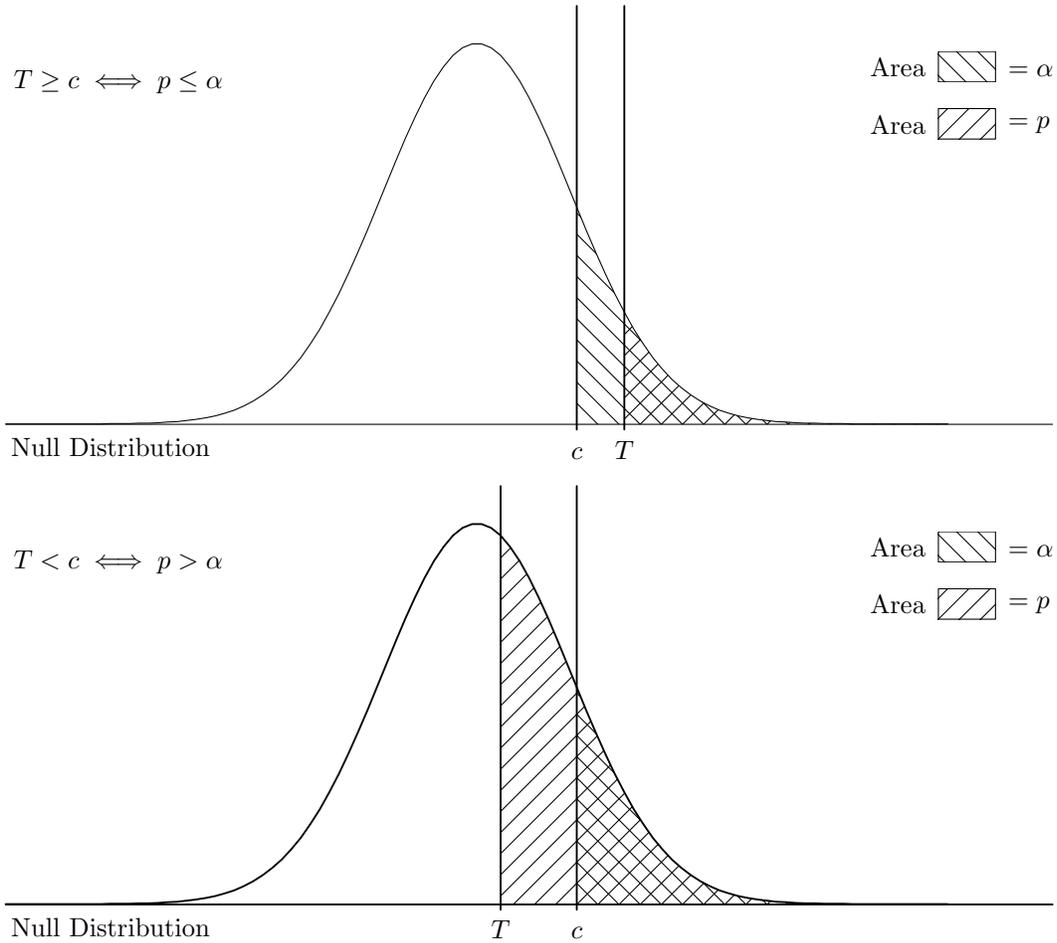}
\caption[]{\label{fig::pval} 
Schematic illustration of the relationship between the p-value $p$ and the Type I error probability $\alpha$
and the equivalent relationship between the test statistic $T$ and the critical threshold $c$.
The curves in both panels represent the distribution of the test statistic under the null hypothesis.
In the top panel, the test statistic $T$ is bigger than the critical threshold $c$ 
(leading to the null hypothesis being rejected).
In the bottom panel, the test statistic $T$ is smaller than the critical threshold $c$
(leading to the null hypothesis being maintained).
The p-value corresponds to the area under the curve and to the right of $T$;
this area is hatched /// in the figure.
The Type I error probability $\alpha$ corresponds to the area under the curve and to the
right of $c$;
this are is hatched $\backslash\backslash\backslash$ in the figure.}
\end{figure}

To clarify these ideas,
we begin by considering a test of the null hypothesis at a single pixel in the source detection problem.
It is common practice to choose a critical threshold that maximizes the power subject to capping the
probability of a false detection (i.e., Type I error) at a pre-specified \emph{significance level} $\alpha$.  
For example, 
if the test statistic has an approximately Gaussian distribution then choosing
$\alpha = 0.05$ is equivalent to using a ``$2\sigma$'' threshold.  
See Figure \ref{fig::pval}
for more on the relationship between $\alpha$ and the critical threshold.
A useful quantity to compute is the p-value.
The p-value is defined as the probability \emph{when the null hypothesis is true} of getting a test
statistic (e.g., normalized photon count) that is at least as extreme as the observed test
statistic.  
As illustrated in Figure \ref{fig::pval},
there are two equivalent ways to decide whether to reject
the null hypothesis: reject when the test statistic is bigger than the
critical threshold or when the p-value is less than $\alpha$.

Next, consider how the situation changes when we test the null hypotheses at many different pixels simultaneously.
We again select a critical threshold, or equivalently a significance level,
and this threshold is applied as above at each pixel to reject or maintain the null hypothesis.
However, because there are many tests being performed,
there are many more ways to make errors.
The following are commonly used approaches to multiple hypothesis testing,
to which we will later compare the FDR method.
\begin{itemize}
\item Naive Multiple Testing.
Use the same threshold (e.g. $2\sigma$) as is used for a single test.
\item 3$\sigma$ Multiple Testing.
Increase the threshold to something more stringent than
naive thresholding, such as a 3$\sigma$ cutoff,
using the same threshold regardless of the number of tests.
\item The Bonferroni Method.
Reject any hypothesis whose p-value is less than
$\alpha^\prime_N = \alpha/N$, where $N$ is the number of tests.
\end{itemize}
Naive multiple testing
rejects any pixel whose p-value
is less than the $\alpha$ defined as for a single test.
Unfortunately, this leads to a higher than expected rate of making Type I errors:
the chance of at least one false rejection over the many pixels
is much larger than $\alpha$.
For example, if $\alpha = 0.05$ and there are 10,000 pixels,
then the chance of at least one false discovery is 
$1 - (1 - 0.05)^{10,000} \approx 1 >> \alpha$.

This naturally leads to the idea of increasing the critical threshold (i.e., $\sigma$-cutoff)
to make the test at each pixel more stringent.
As shown in Figure \ref{fig::pval}, 
increasing the threshold is equivalent to replacing the significance level $\alpha$ by a smaller number $\alpha^\prime$.
For example, choosing $\alpha' = 0.01$ corresponds to using a $3\sigma$ cutoff.
(Because this choice is so common in astrophysical data analysis,
we call this ``$3\sigma$'' multiple testing.)
While this adjustment does reduce the average number of false discoveries relative to naive multiple testing,
it has the same problem: the probability of making a false discovery is typically much larger than the 
desired significance level $\alpha$.
The difference is that this probability grows more slowly with the number of tests.
For example, with $\alpha = 0.05$ and $\alpha' = 0.01$ and $N = 100$,
the probability of at least one false discovery is $1 - (1 - 0.05)^{100} \approx 0.9941$ for naive multiple testing
and is $1 - (1 - 0.01)^{100} \approx 0.6340$ for $3\sigma$ multiple testing.
When $N = 1000$, both probabilities are larger than $0.9999$.

The same argument applies to any critical threshold that is fixed relative to the number of tests $N$
and suggests that one should increase the threshold with $N$.
This is equivalent to choosing a significance level $\alpha'_N < \alpha$ as a function of $N$.
The Bonferroni method takes $\alpha'_N = \alpha/N$. It can be shown that Bonferroni
guarantees that the probability of making at least one false rejection is no greater than
$\alpha$.  Unfortunately, this control on false rejections comes at a high cost:
the probability of erroneously maintaining the null hypothesis
goes to one as $N$ gets large.

To summarize, in the Naive ($2\sigma$) multiple testing,
we allow many false discoveries in return for more correct detections.
In the Bonferroni method,
we tightly control the propensity for making false discoveries but as a result
tend to miss many real detections.
These two methods represent the opposite extremes.
Naive and $3\sigma$ multiple testing are the typical methods of choice in astrophysical data analysis,
but although the latter is more stringent,
both suffer the same fate when the number of tests is large.
With the vast data sets being acquired today, 
this is a common and potentially severe problem.
Methods that control the False Discovery Rate, described in the next section,
are intermediate between these extremes.
These methods adapt to the size of the data to
give control of false discoveries comparable to the fixed threshold methods
while maintaining good power.

\section{The FDR Method} \label{sec::fdr}

The solution that we put forward in this paper is the False Discovery Rate
(FDR) method due to Benjamini \& Hochberg (1995).  FDR improves on existing
methods for multiple testing: it has higher power than Bonferroni, it controls
errors better than the naive method, and it is more adaptive than the
$3\sigma$ method.  Moreover, the FDR method controls a measure of error that
is more scientifically relevant than other multiple testing procedures.
Specifically, naive testing controls the fraction of errors \emph{among those
tests for which the null hypothesis is true} whereas FDR controls the fraction of errors
\emph{among those tests for which the null hypothesis is rejected}.
Since we do not know \emph{a priori} the number of true null hypotheses
but we do know the number of rejections,
the latter is easier to understand and evaluate.
 
Suppose we perform $N$ hypothesis tests.
We can classify these tests into four categories as follows,
according to whether
the null hypothesis is rejected and whether the null hypothesis is true:
$$
\vbox{
\begin{center}
\begin{tabular}{l|ll|l} 
            & Reject Null & Maintain Null&  \\ \hline
Null True   & $\N1|0$     & $\N0|0$    & $\N.|0$     \\ \hline
Null False  & $\N1|1$     & $\N0|1$    & $\N.|1$     \\ \hline
            & $\N1|.$     & $\N0|.$    & $N$         \\
\end{tabular}
\end{center}
\begin{center}
Table 1. Summary of outcomes in multiple testing.
\end{center}
}
$$
The columns in Table 1 are the results of our testing procedure (in which
we either maintain or reject a test).  The rows in Table 1 are the true
numbers of source (null false) and background (null true) pixels.
For example, $\N1|0$ is the number of false discoveries,
$\N1|1$ is the number
of correct discoveries, $\N0|0$
is the number of correctly maintained
hypotheses, and $\N0|1$
is the number of falsely maintained hypotheses.  We
thus define the false discovery rate FDR to be
$$
\FDR = \frac{\N1|0}{\N1|.} = 
\frac{\N1|0}{\N1|0 + \N1|1},
$$
where FDR is taken to be 0 if
there are no rejections.
This is the fraction of rejected hypotheses
that are false discoveries.
In contrast to Bonferroni, which
seeks to control the chance of even a single false
discovery among \emph{all} the tests performed,
the FDR method controls the proportion of errors among
those tests whose null hypotheses were rejected.
Thus, FDR attains higher power
by controlling the most relevant errors.

We first select an $0\le \alpha\le 1$.
The FDR procedure described below
guarantees\footnote{
More precisely,
the procedure makes the stronger guarantee that
$\langle\FDR\rangle \le \alpha\cdot \N.|0/N$, where the right-hand
side is always $\le \alpha$.
If the test statistic has a continuous distribution, the first inequality is also
an equality.} that
\begin{equation}\label{eq::fdr}
\langle \FDR \rangle \le \alpha.
\end{equation}
In contrast,
the naive and $3\sigma$ multiple testing procedures guarantee that 
$$
\langle {\rm PFD}\rangle = \alpha,
$$
where PFD (``Proportion of False Discoveries'')
equals $\N1|0/N$
and
where, for instance, 
$\alpha = 0.05$ for a $2\sigma$ cutoff and $\alpha = 0.01$ for a $3\sigma$ cutoff.
Similarly, the Bonferroni method guarantees that
$$
\langle{\rm AFD}\rangle \le \alpha
$$
where ${\rm AFD}$ (``Any False Discoveries?'')
equals 1 if $\N1|0 > 0$ and 0 if $\N1|0 = 0$.
The expectations $\langle\cdots \rangle$ in all these expressions 
represent ensemble averages over replications of the data.
Note that, like any statistical procedure,
all these methods control an ensemble average;
they do not guarantee that the realized value is less than $\alpha$ on any one data analysis.

The FDR procedure is as follows.
We first select an $0\le \alpha\le 1$.
Let $P_1, \ldots, P_N$ denote the p-values
from the $N$ tests,
listed from smallest to largest.
Let
$$
d =\max\left\{ j :\ P_j < \frac{j \alpha}{c_N N} \right\},
$$
where $c_N$ is a constant defined below.
Now reject all hypotheses whose 
p-values are less than or equal to $P_d$.
Note that every null hypothesis with $P_j$ less than $P_d$ is rejected
even if $P_j$ is not less than $j\alpha/(c_N N)$.
Graphically, this procedure corresponds to
plotting the $P_j$s versus $j/n$,
superimposing the line through the origin of slope $\alpha/c_N$,
and finding the last point at which $P_j$ falls below the line.

When the p-values
are based on statistically independent tests, 
we take $c_N =1$.
If the tests are dependent, 
we take 
$$
c_N = \sum_{i=1}^N \frac{1}{i}.
$$
Note that in the dependent case,
$c_N$ only increases logarithmically with the number of tests.

The fact that this procedure guarantees that (Eqn. \ref{eq::fdr}) holds is not
obvious. For the somewhat technical proof, the reader is referred to Benjamini
\& Hochberg (1995) and Benjamini \& Yekutieli (1999).  
We give a heuristic argument for a special case in Appendix \ref{sec::fdrproof}.
We also provide a simple step-by-step tutorial and sample code
that may be easily implemented in Appendix \ref{sec::wexamp}

\section{Simulation Comparison of FDR to Other Methods} \label{sec::sims}

Consider a stylized version of the source detection problem with a 1,000 by
1,000 image where the measurement at each pixel follows a Gaussian
distribution.  For simplicity, we assume here that each source is a single
pixel and thus the pixels are uncorrelated, though we return to this
issue in Section \ref{sec::hubbledf}.  We assume that the distribution for
background pixels has a mean $\mu_{back} = 1000$
and a standard deviation $\sigma_{back} = 300$, and that
the distribution for source pixels has a mean $\mu_{source} = 2000$
and a standard deviation
$\sigma_{source} = 1000$.  We perform a test at each of the $N=1,000,000$ pixels.  We use
960,000 background pixels and 40,000 source pixels.  The null hypothesis for
pixel $i$ is that it is a background pixel; the alternative hypothesis for
pixel $i$ is that it is a source pixel.  
The p-value for pixel $i$ is the probability that a background pixel
will have intensity $I^i$ or greater:
$$
\displaystyle
p^i_{val} = \int_{I^i}^{\infty} \frac{1}{\sigma_{back}\sqrt{2\pi}}\,e^{-(I - \mu_{back})^2/2\sigma^2_{back}} dI.
$$
We repeated the simulation 100 times and taking the average counts.
In Table 2, we present our results when
trying to recover the original sources in uncorrelated noise.  Notice that the
$2\sigma$ technique produces 53100 events (or discoveries),
of which 40\% are in reality false.
While the Bonferroni technique produces zero false discoveries,
it only finds 27138 sources out of 40,000 in the image.
The FDR technique yields nearly as many real source detections as the $2\sigma$ technique,
and only 1505 false discoveries, a factor of 15 fewer than $2\sigma$.
We stress here that this advantage over the $2\sigma$ technique is a direct
result of the adaptive nature of FDR and comes at no cost to the user. This
example illustrates the worth of FDR in helping the statistical discovery in
astrophysics, and we thus champion its use throughout the community.

\begin{deluxetable}{cccccc}
\tablenum{2}
\tablewidth{0pt}
\tablecaption{\bf 40000 Pixel Sources in a $1000\times1000$ Pixel Uncorrelated Gaussian Background }
\tablehead{
\colhead{\vtop{\hbox{\enspace Method}\vskip 4pt\hbox{\strut($\alpha = 0.05$)}}} & \colhead{\vtop{\hbox{p-value}\vskip 4pt\hbox{\strut\enspace cutoff}}} & 
\colhead{$\left\langle\N1|1\right\rangle$} & \colhead{$\left\langle\N1|0\right\rangle$} &
\colhead{$\left\langle\N0|1\right\rangle$} & \colhead{$\left\langle\N0|0\right\rangle$}}
\startdata
$\langle {\rm FDR} \rangle$  & $1.6~~\,\times10^{-3}$ & 30389  & \phantom{2}1505 & ~9611 & 958495    \nl
2$\sigma$ &  $2.275\!\times10^{-2}$  & 31497 & 22728 & ~8503 & 937272 \nl
$\alpha/N$ & $5.0~~\,\times10^{-8}$  & 27137 & \phantom{2272}0 & 12863 & 960000 \nl
\enddata
\end{deluxetable}

To show how the FDR procedure adapts, we perform this same simulation using
different alternative (or source) distributions.  In particular, we increase
the mean intensity of the source pixels over a range from 1500 to 3000 counts
per pixel, that is from values that are close to the background (sky) mean to
values that are well separated from the background (sky) mean.  Figure
\ref{fig::fdrMonte}, displays the results of four such simulations.  Notice
that the FDR method adapts to the data: the p-value cutoff changes
systematically as the source intensity changes.  Also notice in this figure
that the recovery rate of FDR is nearly that of the much more liberal
$2\sigma$ method, and always greater than the ultra- conservative Bonferroni
technique.  The average false discovery rate over the simulation is $\le 0.05$
for the FDR method, as it should be, and is zero for Bonferroni.  The false
discovery rate for the $2\sigma$ method varies widely but never gets below
35\%, even when the source and background distributions are entirely
separated.  We stress again that this problem can not be solved by simply
increasing the threshold to, say, $3\sigma$ since one then trades a low false
rejection rate for power in detecting real rejections.

\section{Other Examples of FDR in Astrophysics} \label{sec::phys-examp}
In this section, we show two applications of the FDR procedure to astrophysical
data. Specifically, we consider the one-dimensional power spectrum of the
distribution of galaxies and clusters in the universe and source detection in
a two-dimensional image of the sky.  However, FDR can be applied to any
problem involving multiple hypothesis testing.

\subsection{Detecting Features in the Matter Power Spectrum}
During its first $\simeq100,000$ years, the Universe was a fully ionized
plasma with a tight coupling between the photons and matter via Thompson
scattering. A direct consequence of this coupling is the acoustic oscillation
of both the primordial temperature and density fluctuations (within the
horizon) caused by the trade--off between gravitational collapse and photon
pressure.  The relics of these acoustic oscillations are predicted to be
visible in both the matter and radiation distributions, with their relative
amplitudes and locations providing a powerful constraint on the cosmological
parameters ({\it e.g.}  $\Omega_{total}, \Omega_{vacuum}, \Omega_{baryons},
\Omega_{dark~matter}$).  Recently, the BOOMERANG and MAXIMA teams announced
the first high confidence detection of these acoustic oscillations in the
temperature power spectrum of the Cosmic Microwave Background (CMB) radiation
(Netterfield et al. 2001, Lee et al. 2001).  Recently, Miller, Nichol, \&
Batuski (2001) used FDR to show that the corresponding acoustic oscillations
are in the matter power spectra of three independent large-scale structure
datasets.

In Figure \ref{fig::baryons} (left) we plot $P(k)$, the power spectrum, for
the three large-scale structure samples described in Miller et al. 2001.
The solid line is the smooth,
featureless, null-hypothesis as discussed in Miller et al. (2001).  The
circled points are rejected using the FDR procedure with $\alpha = 0.25$,
while the points outlined with squares are rejected with $\alpha=0.10$.  We
detect the ``valleys'' at both $k \sim 0.035h$Mpc$^{-1}$ and at $k \sim
0.09h$Mpc$^{-1}$.  With $\alpha = 0.1$, we only reject  five point, while
with $\alpha =0.25$ we reject eight points.
The properties of the FDR bound suggest the
fluctuations are true outliers against a smooth, featureless spectrum.
Note that each of the three data sets contributes to the features,
and so the detection is not dominated by one sample.  In the right panel, we
plot the p-value versus index for the combined LSS dataset.  The solid red line
is for $\alpha = 0.25$, and anything to the left of the vertical blue dotted
line is detected as being part of a feature in the power spectrum with a
maximum false discovery rate of 0.25.

As discussed in Miller et al. 2001, the use of FDR improves our ability to 
detect these features in these three power spectra. This would not have been
as convincing if we had used the usual ``$2\sigma$'' procedure of multiple
hypothesis testing {\it i.e.}  demanding that all the points in the
``valleys'' in the $P(k)$ (Figure \ref{fig::baryons}) be greater than $2\sigma$
from a smooth function.
%% ATTN: I don't like the following but include it if you want to
%This clearly illustrates the power of new statistical
%tools in enhancing the analysis of astronomical data sets thus allowing new
%discoveries from data commonly believed to be ``too noisy''.

\subsection{Source Detection in the Hubble Deep Field}\label{sec::hubbledf}

In Section \ref{sec::sims},
we showed the results of using FDR on a simulated
image of the sky (with sources in Gaussian background). Here, we will utilize
FDR on the image created by Szalay, Connolly, and Szokoly (1999), who applied
a novel source detection algorithm to the Hubble Deep Field (HDF)\footnote {Based on
observations made with the NASA/ESA {\it Hubble Space Telescope}, obtained
from the data archive at the Space Telescope Science Institute, which is
operated by the Association of Universities for Research in Astronomy, under
NASA contract NAS5-26555.}. 
The basis for the Szalay et al. analysis is straightforward:
for an uncorrelated set of images with zero mean and
unit variance (i.e. the normalized sky) the probability distribution of the pixel
values forms a chi-square distribution. 
They combine the
individual pass bands from the HDF into one image used for source pixel detection.
However, instead of summing (with or without weights) the individual colors
images, Szalay et al.  build up a $\chi^2$ (or $y$-image). This $y$-image is
constructed by calculating  $\chi^2$ for $n$ degrees-of-freedom (where $n$ equals the
number of pass-bands). The source pixels are in a pre-determined background
with a well known error in each band.  Each pixel in the $y$-images is then a
$\chi^2$ value with $n$ degrees of freedom.
The chi-square construction provides an almost optimal use of the color
information.
Szalay et
al.  provide some nice examples of how source pixels are ``amplified'' compared
to the individual pass-bands (see their Figure 5).

The next step in Szalay et al. (1999) was to determine the threshold in
$\chi^2$ space (probability space) above (below) which a pixel can be
considered statistically above the background. Szalay use the so-called
optimal Bayes Threshold, i.e. the threshold that chooses the class with
maximal posterior probability.
Szalay et al.~determine that the ``best'' threshold is for a $\chi^2$ value
$\ge 13.9129$. If we convert from $\chi^2$ to the probability of randomly
finding a pixel with this $\chi^2$ or higher, we find the p-value equals
$1 - 0.9924 = 0.0076$.  This p-value threshold, which is highlighted in Table 3,
corresponds (in Gaussian terms) to a 2.42$\sigma$ detection per pixel.  In
Table 3, we show how varying degrees of $\alpha$ correspond to the false
discovery rate using the methods described in this paper. In other words,
we can use the FDR procedure to place bounds on the false discovery rate
for past analyses that used standard thresholding techniques. Note in Table 3 that
the cut-off used by Szalay et al. corresponds to a false discovery of
11\%. This means that, of the total number of source pixels used to find
galaxies in the $y$-image, 11\% could be in error (on average). Therefore
the Bayes Threshold that they applied did not introduce a large amount
of false discoveries. In
Figure \ref{fig::hdf}, we plot the rejected pixels for two different
$\alpha$'s in a piece of the $y$-image.

The above example shows how the FDR procedure provides a much more
relevant (and useful) quantity: a bound on the fraction of false
discoveries out of the total number of source pixel detections. If the
derived false discovery rate had been extremely high (say 50\%), than
their Bayes Threshold used would have been too low, and their conclusions 
less reliable. However, this was not the case. Unfortunately, 
standard thresholding techniques do not provide this information.
In the above example, we assumed that the individual pixels are
uncorrelated. This is not entirely true, since the point spread function for
the HST will result in correlations between blocks of 10 or so pixels.
Also, true sources (e.g. galaxies) must have a minimum number of
connected source pixels (which are physically correlated).
We could also use FDR with the factor ($c_N$) for correlations included.
In a future work (see e.g. Hopkins et al. in preparation),
we will examine in greater detail the process of accounting for these correlations
in the FDR procedure.

\begin{deluxetable}{cccc}
\tablenum{3}
\tablewidth{0pt}
\tablecaption{\bf Using FDR on the $\chi^2$ Image of Szalay et al. 1999}
\tablehead{
\colhead{$\alpha'$} & \colhead{\mbox{p-cutoff}} & \colhead{Detection Threshold} &
\colhead{Number of rejected pixels}}
\startdata
0.150 & 0.01172 & 2.27$\sigma$ & 78122 \nl
{\bf 0.108} & {\bf 0.00761} & {\bf 2.42$\sigma$} &{\bf 70444} \nl
0.100 & 0.00688 & 2.46$\sigma$ & 68882 \nl
0.050 & 0.00294 & 2.75$\sigma$ & 58598 \nl
0.010 & 0.00047 & 3.31$\sigma$ & 46878 \nl
\enddata
\end{deluxetable}

\section{Discussion} \label{sec::disc}

The methods discussed in this paper are frequentist (non-Bayesian) methods.
Readers who are familiar with Bayesian inference might ask
if there are Bayesian analogs of the FDR procedure.
Indeed, one can construct a Bayesian
version of the FDR procedure.
For example,
let $H_i=0$ mean that the null hypothesis holds
for the $i^{\rm th}$ hypothesis and let
$H_i=1$ mean that the alternative hypothesis holds
for the $i^{\rm th}$ hypothesis.
Let ${\bf p}$ denote the vector of p-values.
By Bayes' theorem,
the posterior probability that $H_i=1$ is
$$
Pr(H_i =1|{\bf p})=
\int \frac{f_1(p;\theta)\pi}{f_1(p;\theta)\pi + f_0(p)(1-\pi)} 
g(\pi, \theta| {\bf p}) d \pi d\theta
$$
where
$f_0(p)\equiv 1$ is the density of p-values under the null hypothesis
and $f_1(p;\theta)$ is the density of p-values under the alternative hypothesis
(indexed by a vector of parameters $\theta$),
$\pi = Pr(H_i=1)$ is the prior probability that
a null hypothesis is false and
$g(\pi, \theta| {\bf p})$ is the
posterior for the parameters
$\pi$ and $\theta$
given by
$$
g(\pi,\theta|{\bf p}) =
\int\int 
\sum_{S \in {\cal S}} \prod_i 
\left( \pi f_1(p_i; \theta) \right)^{S_i}
\left( (1-\pi) f_0(p_i) \right)^{1-S_i}
w(\pi,\theta) d\pi d\theta
$$
where the sum is over all
vectors $S$ of length $n$
consisting of 0's and 1's and
$w(\pi,\theta)$ is a prior on $\pi$ and $\theta$.
(We have expressed this in terms of p-values to
be consistent with the rest of the paper but
the expressions can be written in terms of the test statistics directly.)
Using similar calculations, one can derive
a Bayesian posterior probability 
on the false discovery rate of a given set of hypotheses.
Developing these ideas further and
comparing them to the non-Bayesian method discussed
in this paper will be the subject of future work.
We will make two points, however.
First,
as always, the Bayesian approach requires more assumptions,
for example, specification of the form of $f(p;\theta)$
and the prior $w(\pi,\theta)$.
Second, in estimation problems, Bayesian and non-Bayesian
methods agree in large samples.
For example, an interval with Bayesian posterior probability 0.95
will cover the true parameter value with
frequency probability 0.95 + $O(n^{-1})$
where $n$ is sample size.
A well known fact in statistics
is that this agreement between Bayes and non-Bayes methods
breaks down in hypothesis testing situations.
Indeed, this is the source of much debate in statistics.
This is not meant to speak for or against the Bayesian approach
but merely to warn the reader that there are
subtle issues to consider (see Genovese and Wasserman (2001) for
more details). 

\section{Conclusion}

In this paper, we have introduced the False Discovery Rate (FDR) to the
astronomical community. We believe this method of performing multiple
hypothesis testing on data is superior to the ``$2\sigma$'' method
commonly used within astrophysics.
We say this for two reasons: {\it i)} FDR is adaptive to
the number of tests performed and thus provides the same power as the usual
``$2\sigma$'' test, but with a greatly reduced false rejection rate (see the
simulations in Section 5); {\it ii)} Correlated errors are easy to incorporate
into FDR which is of great benefit to many astronomical analyses. To help aid
the reader, we have provided several examples of how to use FDR 
and a tutorial on how to implement FDR within data analyses. We hope
that the astronomical community will use this powerful new tool.

The authors would like to thank Eric Gawiser  for his comments and suggestions. This work was support in part by NSF KDI grant XXXXXXX and by NSF grant SES-9866147.

\newpage

% references

\appendix
\section{A Heuristic Proof of FDR} \label{sec::fdrproof}
In this Appendix, we provide an illustrative proof of the False Discovery
Rate procedure.  This heuristic proof is not intended to replace
the full treatment given in Benjamini and Hochberg (1995). We consider
one very simple case with non-overlapping source and null
distributions.
We can compare this simple scenario
to Fig. \ref{fig::fdrMonte}.
We dictate two distinct intensity distributions, $I^{back}$ and
$I^{source}$ for the background and source pixels respectively. The differential
distributions are shown in Fig. \ref{fig::dist}.   The distribution of
source intensities is entirely separated from those of the background.
Let $F$ be the cumulative distribution function for the background intensity.
The p-values are then $ 1 - F(T)$ for a given test statistic, $T$.
Since the intensities of the sources
are so high compared to the background, $p^i_{val} \approx 0 \equiv 0$ for the
$i$ sources.  

What we would like to know is the distribution of $F(T)$ for 
many realizations of the test statistic, $T$.
Let $Y$ denote $F(T)$, which is a random variable
between zero and one (since $F$ is a cumulative distribution).
Now, consider
the cumulative distribution function for random variable $Y$:
\begin{equation}
G(u) = P \{ Y \le u \}.
\end{equation}
We can derive $G(u)$ for all $u$ by considering the
the following regimes:
\begin{eqnarray}
\qquad\mbox{For~} u < 0 &  G(u) = 0 \\ \nonumber
\qquad\mbox{For~} u > 1 &  G(u) = 1 \\ \nonumber
\qquad\mbox{For~} 0 \le u \le 1 & G(u) = u  
\end{eqnarray}
The final regime needs some explanation. Since $F$ is a
cumulative distribution function, it is monotonically
increasing and continuous.  This means that $F$ has an
inverse function, $F^{-1}$ which is also monotonically
increasing.  We can apply this inverse function in the following way:
\begin{eqnarray}
G(u) & = & P \{Y \le u \} \nonumber \\
& = & P \{F(T) \le u \} \nonumber  \\
& = & P \{ F^{-1}(F(T)) \le F^{-1}(u) \} \nonumber \\
& = & P \{ T \le F^{-1}(u) \} \nonumber \\
& = & F(F^{-1}(u)) \nonumber \\
& = & u
\end{eqnarray}
In words, the cumulative distribution function for the random variable $Y$
is between zero and one and increases linearly with slope one. 
The probability density of $Y$ is then $\frac{dG}{du} = 1$ for $0 \le u \le 1$,
and zero elsewhere.
Therefore, the distribution of $Y = F(T)$ is uniform between between zero and one.

In our very simple case (with two entirely separated distributions),
we will perform the FDR procedure by sorting the p-values. Recall
that the p-values for the source distribution are so small that we
take them to be zero. The
p-values for the background are
drawn from the probability density function for random variable $Y = F(T)$.
Since the distribution of $F(T)$ is uniform, the distribution
of each p-value, $p^i_{val}(T)$, is also uniform.
But what will a plot of the $n$ ordered p-values look like?
We need to first determine the density function and
fortunately, there is a theorem in order statistics which states:
\begin{quote}
Let $U_1, \ldots, U_n$ be independent identically distributed
continuous random variables with common distribution function $H(u)$
and common density function $h(u)$. IF $U_{(i)}$ denotes the $i$th-order
statistic, then the density function of $U_{(i)}$ is given by:
\begin{equation} 
g_{(i)}(u_i) = \frac{n!}{(i-1)!(n-i)!}(H(u_i))^{i-1}[1-H(u_i)]^{n-i}h(u_i) 
\end{equation}
\end{quote}

For our case, $H(u_i) =  u_i$, as shown above in equation (A3). If we replace
$H(u_i)$ with $u_i$ in the above equation, we simply have a Beta distribution,
$B(a,b)$ where $a = i$ and $b = n - i + 1$. The Beta distribution
has an expectation value of $\frac{a}{a+b}$ and so:
\begin{equation}
\langle p^i_{val} \rangle = \frac{i}{n+1}
\end{equation}
for the $n$ p-values.

Now, we know the expected values of all of the p-values in this simple example. The
sources have $p^i_{val}=0$ while the background pixels have $p^i_{val}=\frac{i}{n+1}$.
The sorted p-values are shown in Fig. \ref{fig::app_pval}, where we have overplotted
the line $y = \alpha i/N$ where $\alpha = 0.1$.  In this Figure, the
probability below which we reject all pixels (as background), occurs
at the first undercrossing between the line and the p-values (working
from the right). The horizontal line indicates this cutoff probability (p-cutoff).
However, we know that all of the real source pixels 
have p-values set to zero, and so we are rejecting
some pixels that are truly background (those pixels with
$N^{source}/N \le i/N \le N^{reject}/N$). We now wish to show
that the fraction of these mistakenly rejected background pixels
is equal to $\alpha$.  We can do this many ways-- by simply counting
the number of tests with p-values in this region (10) and comparing
to the total number of rejections (100), by
geometrical arguments, or by examining the value of the probability cutoff.
For instance, over a large number of realizations of the intensities, we know that
\begin{equation}
\qquad\mbox{$\langle$ p-cutoff $\rangle$} = \langle \alpha\frac{N^{reject}}{N} \rangle
\end{equation}
but we also know that any p-value is the probability of finding
a background pixel with $I > I^{reject}$. So we may also write:
\begin{equation}
\qquad\mbox{$\langle$ p-cutoff $\rangle$}  = p^{i^{\prime}}_{val} = \langle \frac{N^{reject}_{background}}{N} \rangle
\end{equation}
where $i^{\prime}$ means the index of the p-value below which
we reject all tests as sources.
We can now set equations (A6) and (A7) equal and solve for $\alpha$.
As expected, we
find:
\begin{equation}
\langle \alpha \rangle = \langle \frac{N^{reject}_{background}}{N^{reject}} \rangle.
\end{equation}

\section{A Worked Example} \label{sec::wexamp}

\noindent Suppose we conduct ten tests
leading to the following p-values:

\indent {\verb+pvals+} =  [0.023 0.001 0.018 0.0405 0.006 0.035 0.044 0.046 0.021 0.060].

\noindent Step 1. Sort the p-values:

\indent {\verb+sorted_pvals+} = [0.001 0.006 0.018 0.021 0.023 0.035 0.0405 0.044 0.046 0.060].

\noindent Step 2. Compute
$j\alpha/10$ for $j=1, \ldots 10$ and $\alpha = 0.05$:

\indent {\verb+j_alpha+} = [0.005 0.010 0.015 0.020 0.025 0.030 0.035 0.040 0.045 0.050].

\noindent Step 3.
Subtract the list in Step 1 minus the list in Step 2.

\indent \verb+diff+ = [-0.004 -0.004  0.003  0.001 -0.002  0.005  0.0055  0.004  0.001  0.010].

\noindent Step 4.
Find the largest index (from 1 to 10) for which the
corresponding number in Step 3 is negative.
In this example, it is the 5th p-value $0.0230$
corresponding to the difference $-0.002$.
Note that this step corresponds to finding the largest $j$
such that $P_j < j\alpha /n = 0.05 j /10$.

\noindent Step 5.
Reject all null hypothesis whose p-values
are less than or equal to $0.0230$.
The null hypothesis for the other tests are not rejected.
Thus, in this example, the hypothesis with p-values
0.0010, 0.0060, 0.0180, 0.0210, and 0.0230 are rejected.
The procedure is illustrated in Figure \ref{fig::fdr}. [Note
that the p-values in Figure \ref{fig::fdr} are not the same
as those used in this step by step procedure.]

A seven-step prescription that can be directly applied in
the IDL programming language and analysis package would look like the following:

\verb+IDL>size_pvals=size(pvals); pvals is a vector containing the p-values+

\verb+IDL>sorted_pvals=pvals[sort(pvals)]+

\verb+IDL>alpha=0.05+

\verb&IDL>j_alpha=alpha*(findgen(size_pvals[1])+1)/size_pvals[1]&

\verb+IDL>diff=sorted_pvals-j_alpha+

\verb+IDL>p_cutoff=sorted_pvals[max(where(diff_pval le 0.0))]+

\verb+IDL>events=pvals[(where(pvals le p_cutoff)]+

\noindent
where \verb+events+ is the final vector containing all of the
rejections from the null hypothesis.  These are the ``discoveries'',
and at most $\alpha=5\%$ are false.

%
% Figures
%

\newpage

\begin{figure}[p]
 \plottwo{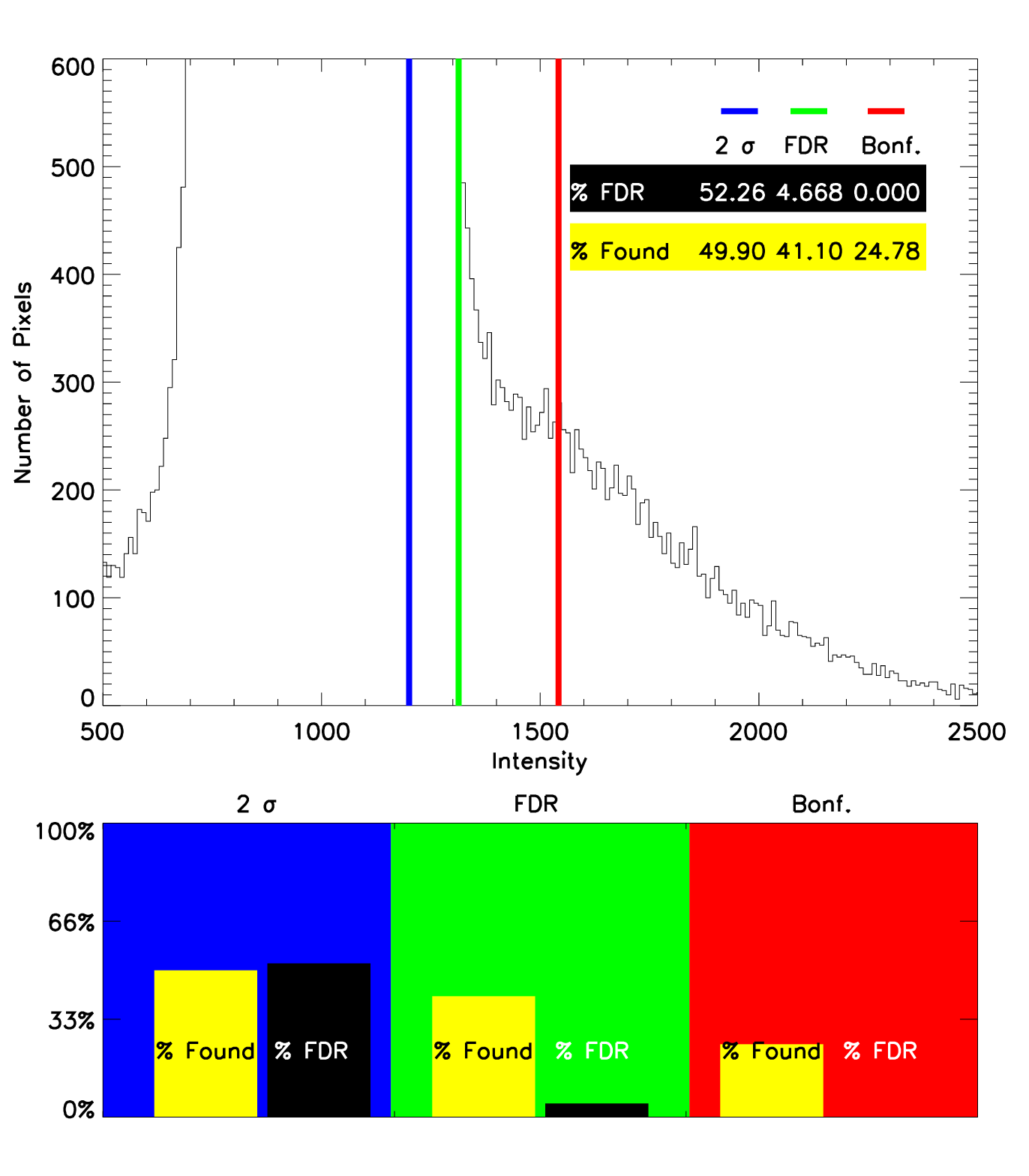}{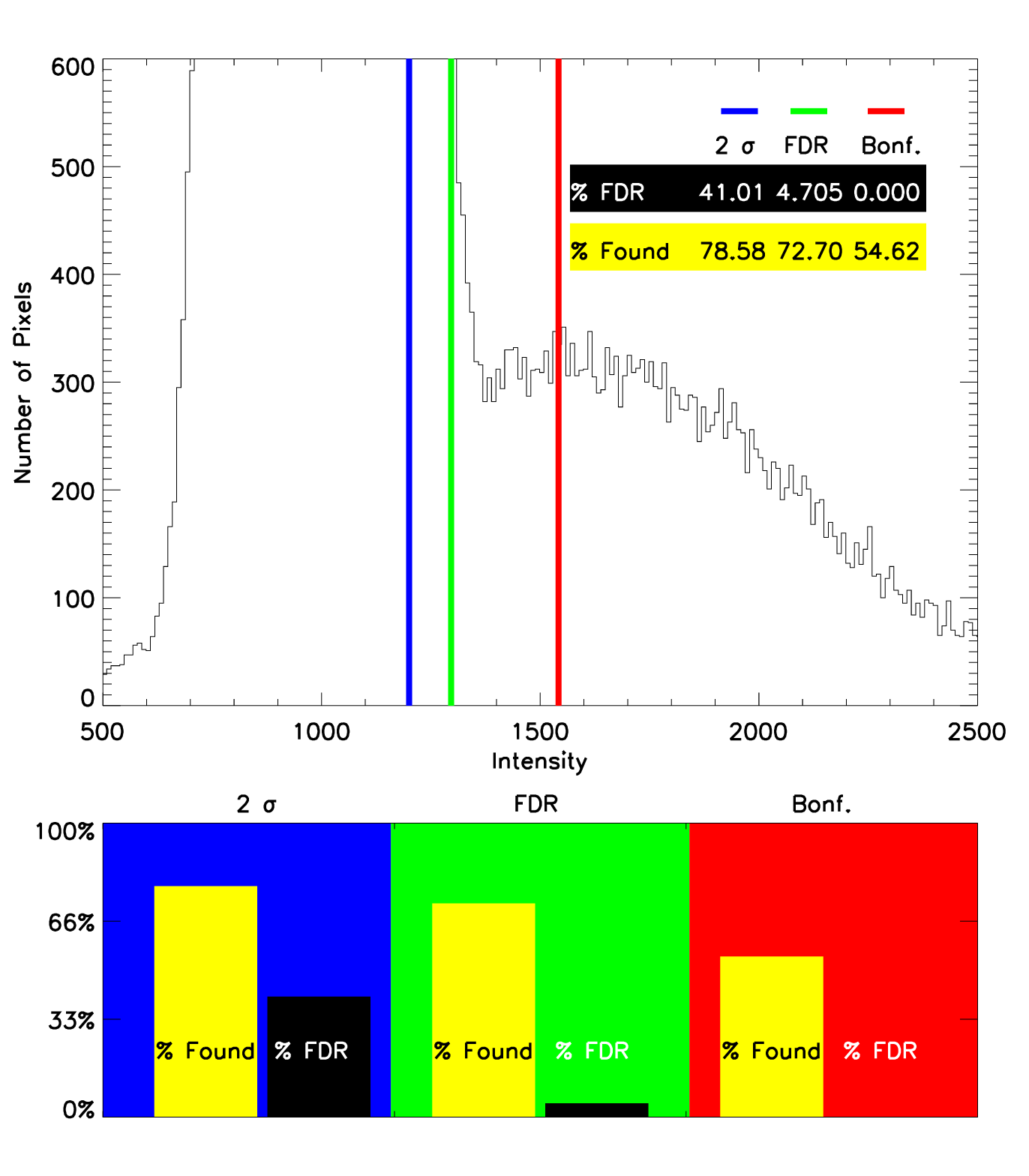}
\end{figure}
\begin{figure}
 \plottwo{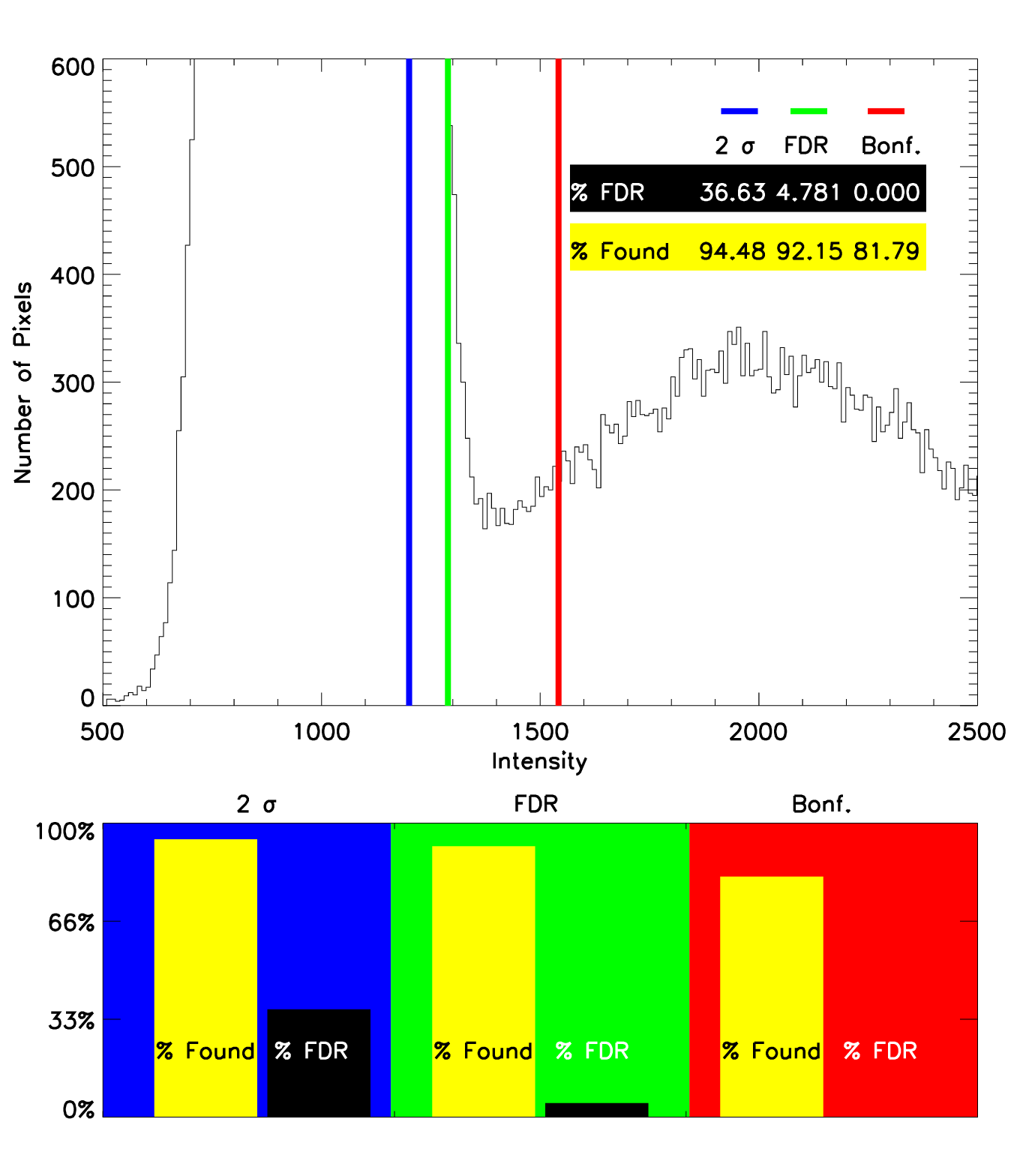}{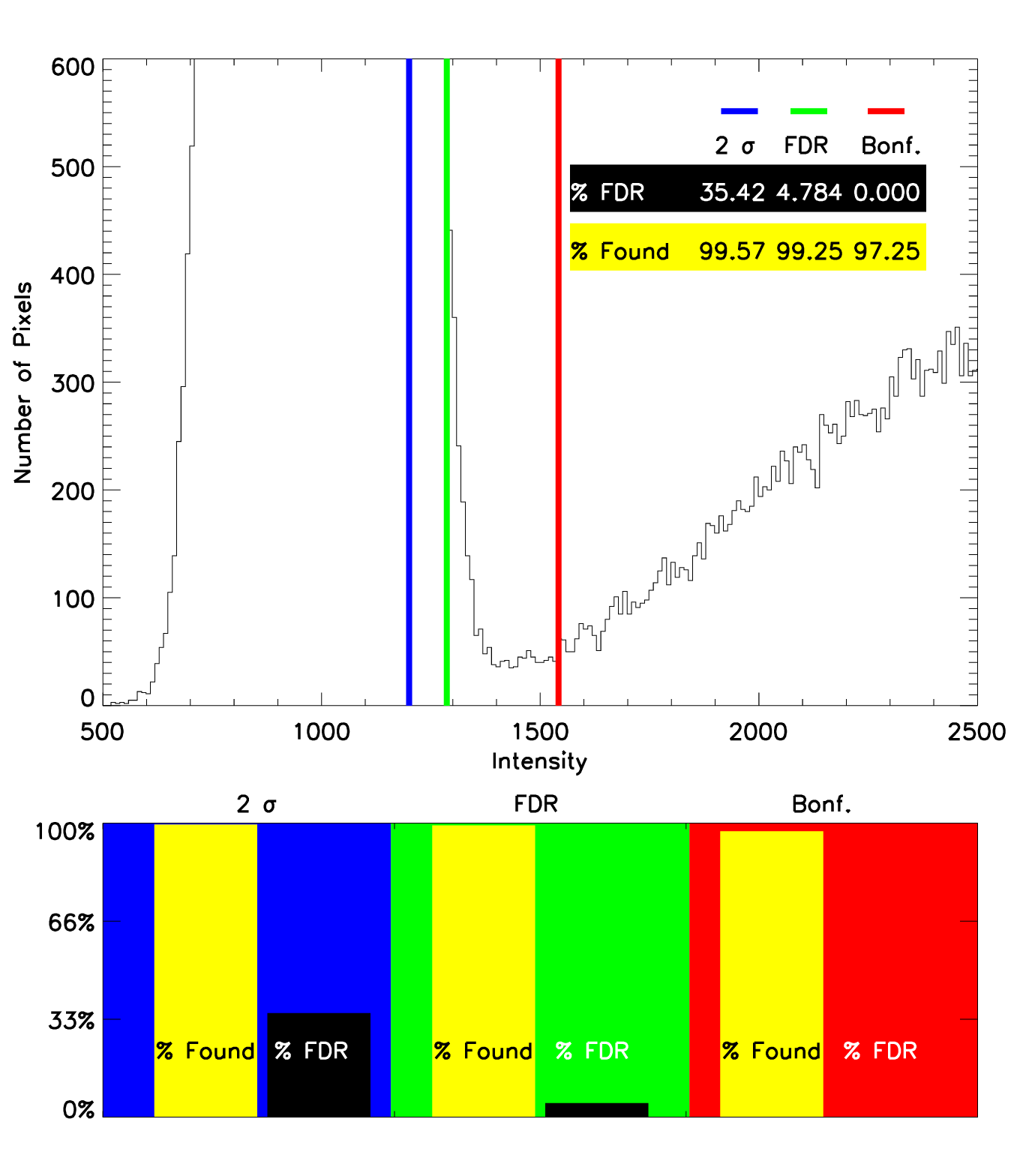}
\caption[]{\label{fig::fdrMonte}
The count distribution for the
960,000 background pixels and 40,000 source pixels. The background 
distribution is Gaussian and peaks off the plot. The four panels
show four different source distributions with different mean intensities
(the dispersion is 1000 in each case). The blue line corresponds to where
a $2\sigma$ cut-off would be made. The green line corresponds to where
the FDR cut-off would be made. The red line corresponds to where the
Bonferroni cut-off would be made. Note that the FDR cut-off moves towards
a lower cut-off as the source and background distributions separate. This
shows the adaptive nature of FDR. Also note that the success rate (\% found)
of FDR nearly matches that of the $2\sigma$ method and is always greater
than the Bonferroni technique. Finally, note that the false discovery rate
remains $< 5\%$ ($\alpha$ was chosen to be 0.05).}
\end{figure}

\newpage

\begin{figure}[p]
\plottwo{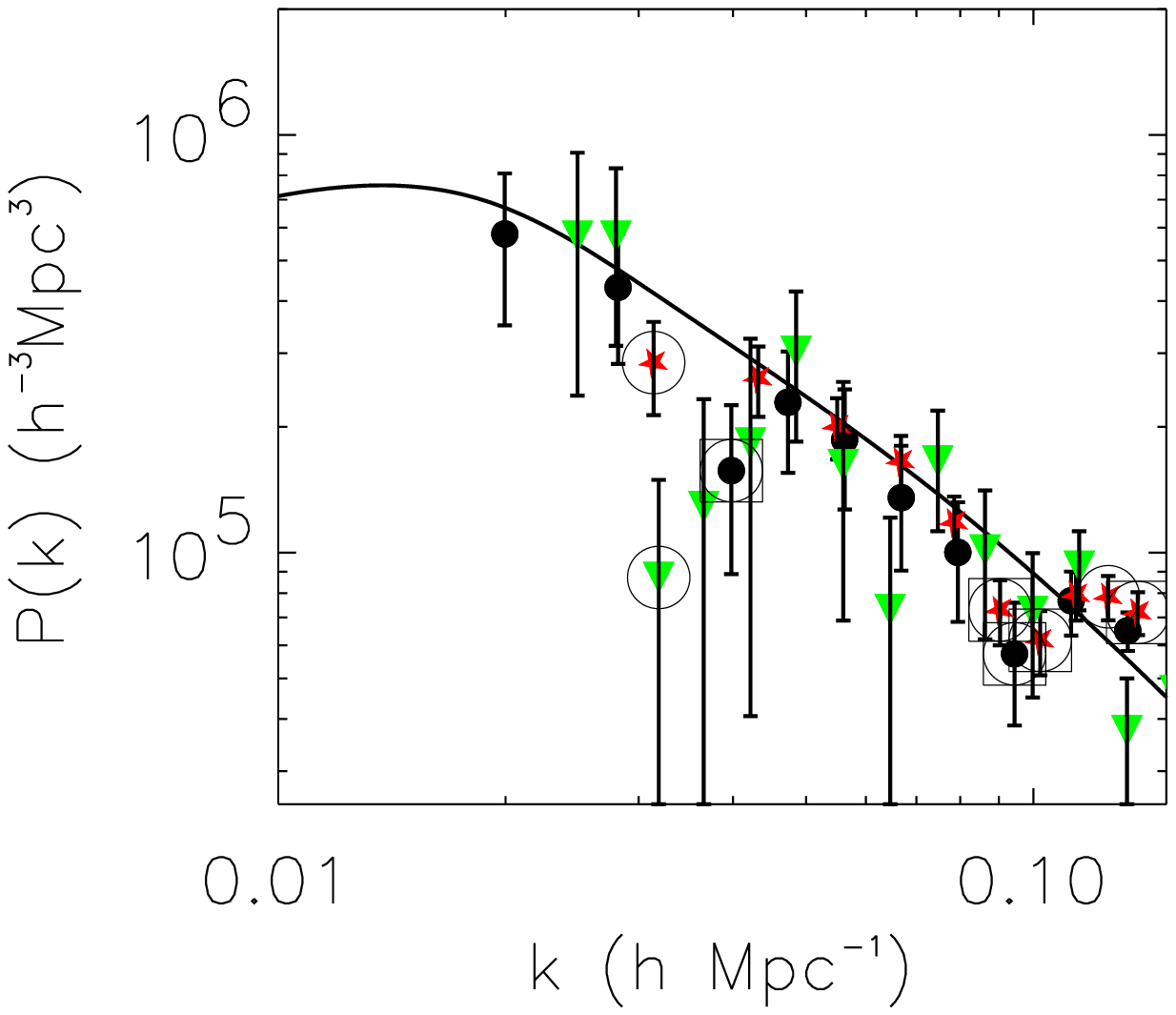}{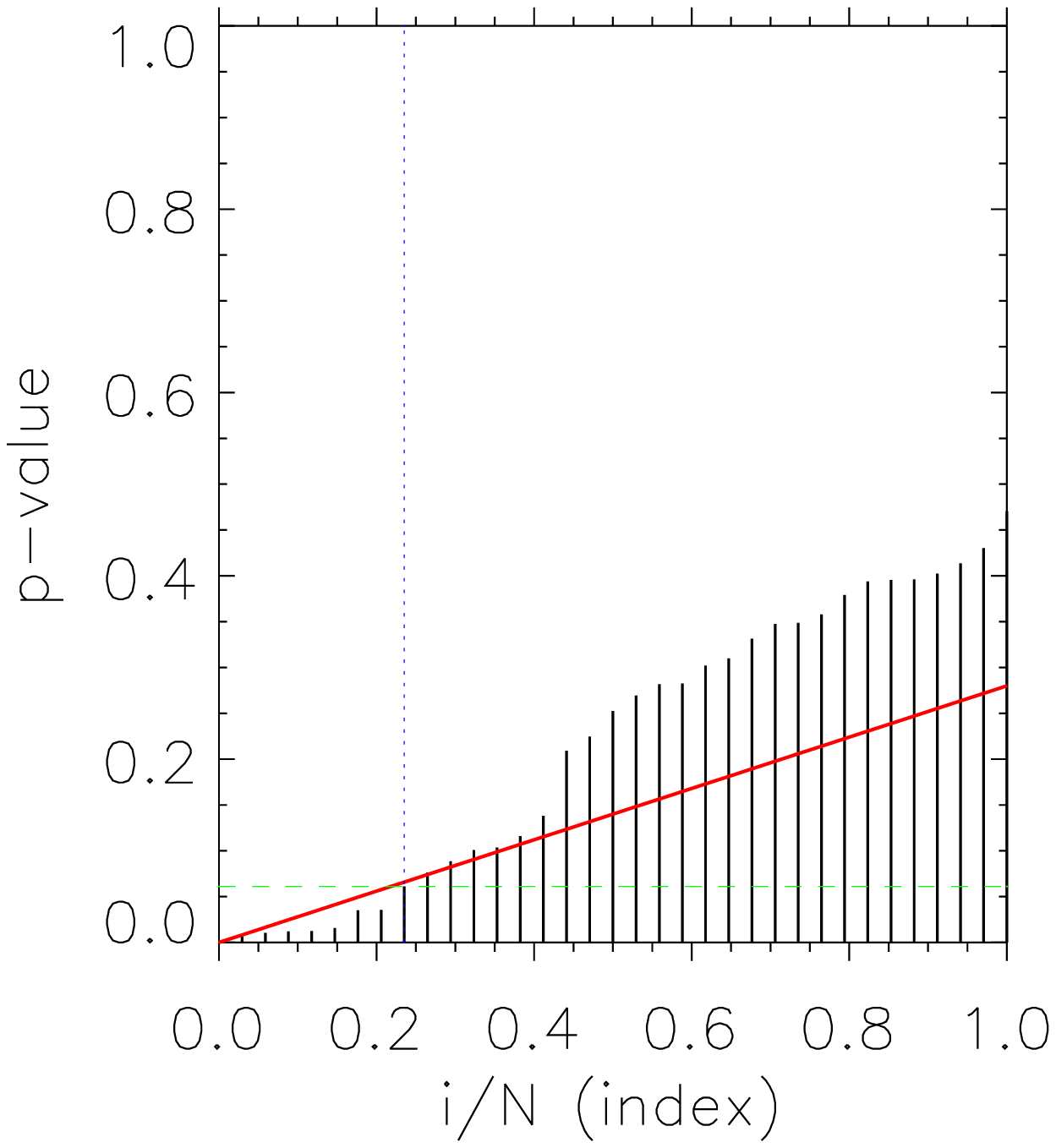}
\caption[]{\label{fig::baryons} {\bf Left}: Here, we show the amplitude shifted power spectra
for the three samples of uncorrelated data. The points highlighted with a circle
denote rejections with $\alpha = 0.25$ (e.g. a quarter of
the rejections may be mistakes). The points highlighted by squares
are for $\alpha = 0.10$ (e.g.  a tenth of the rejections
may be mistakes). The analysis utilizes our best-fit model
with the baryon wiggles removed as the null hypothesis.
By controlling the false discovery rate, we can say with statistical confidence
that the two ``valleys'' are detected as features in the power spectra.
{\bf Right}: The p-values vs. index for the combined data. The red solid
line is for $\alpha = 0.25$, while the horizontal green dashed line shows
the p-cutoff value. Everything to the left of the vertical blue dotted line
is rejected with a false discovery rate of $0.25$.}
\end{figure}

\begin{figure}[p]
 \plottwo{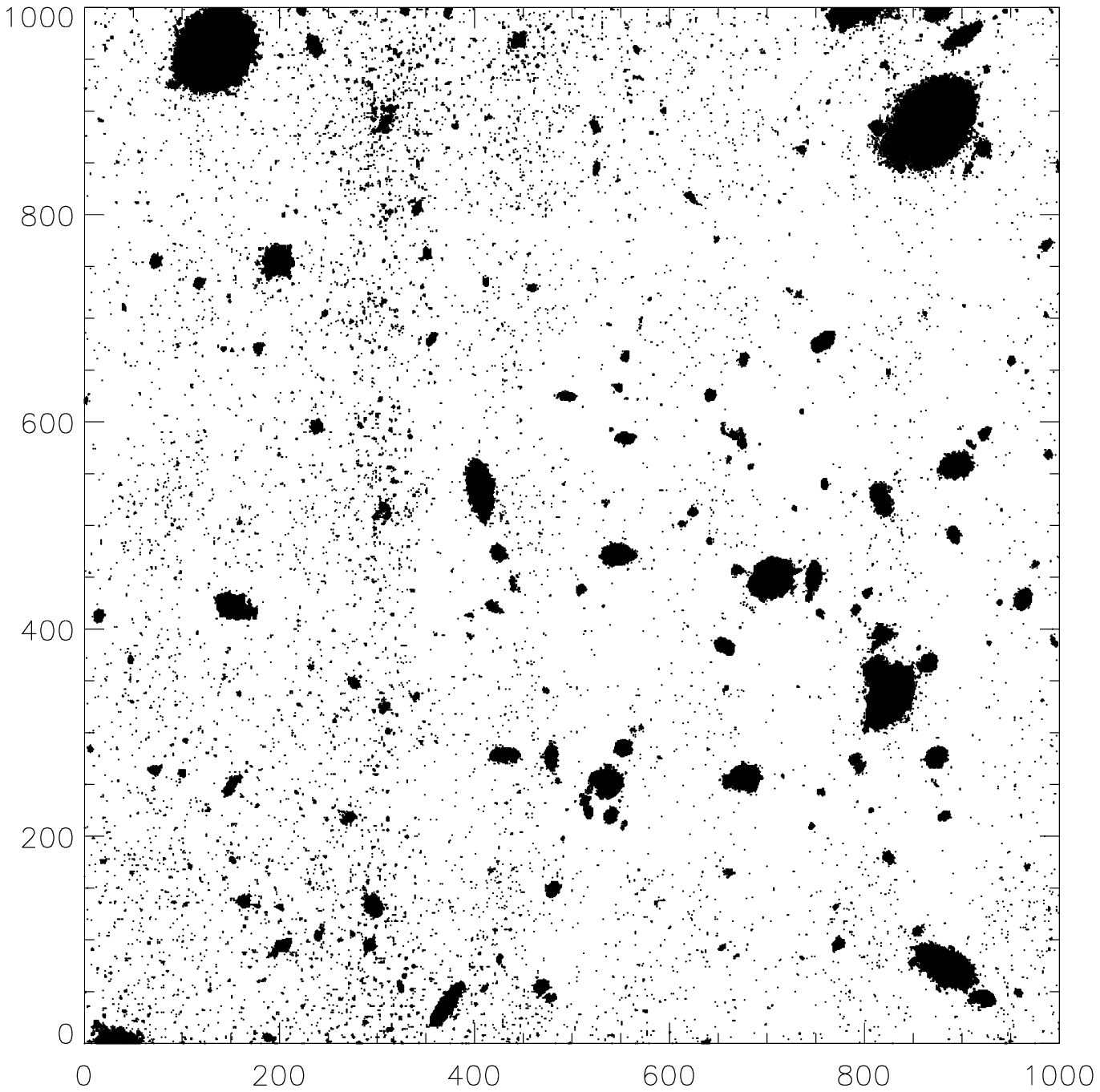}{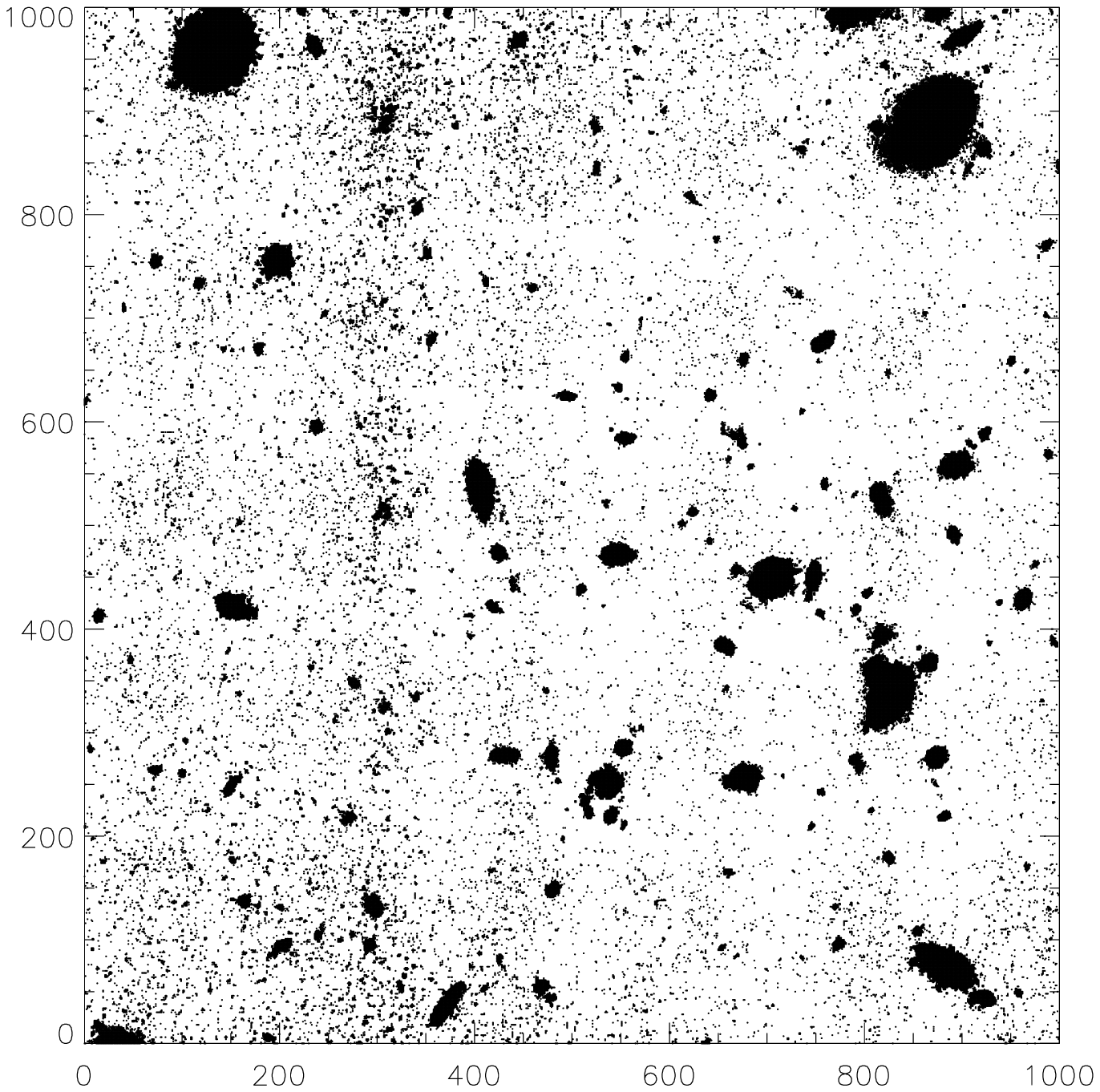}
\caption[]{\label{fig::hdf} 
Rejected pixels from the Hubble Deep Field $\chi^2$ image
of Szalay et al. 1999. Left is for $\alpha = 0.05$ and right is
for $\alpha = 0.15$.}
\end{figure}

\begin{figure}
\plotone{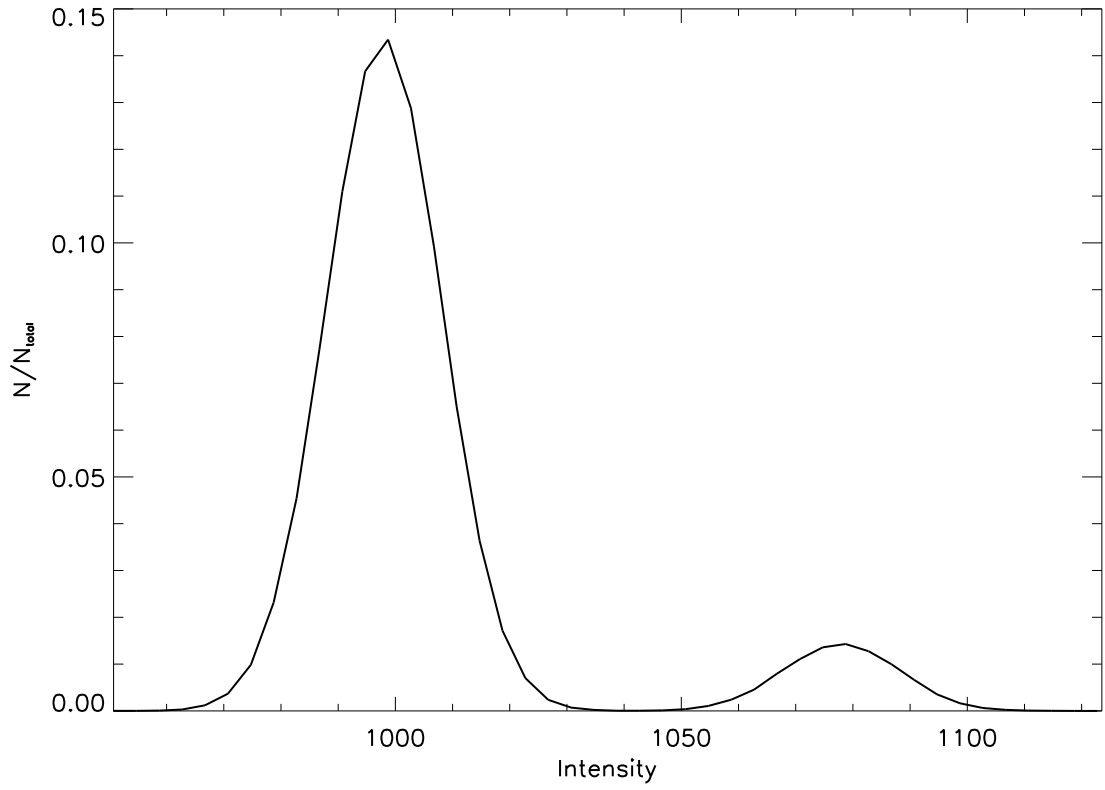}
\caption[]{\label{fig::dist} The differential distribution of pixels. The left
hump is centered on $I = 1000$ and is thus represents
the background distribution. The right hump is centered on
$I = 1080$ and represents the source distribution. 
There are 10 times as many background as source pixels. }
\end{figure}

\begin{figure}[h]
\plotone{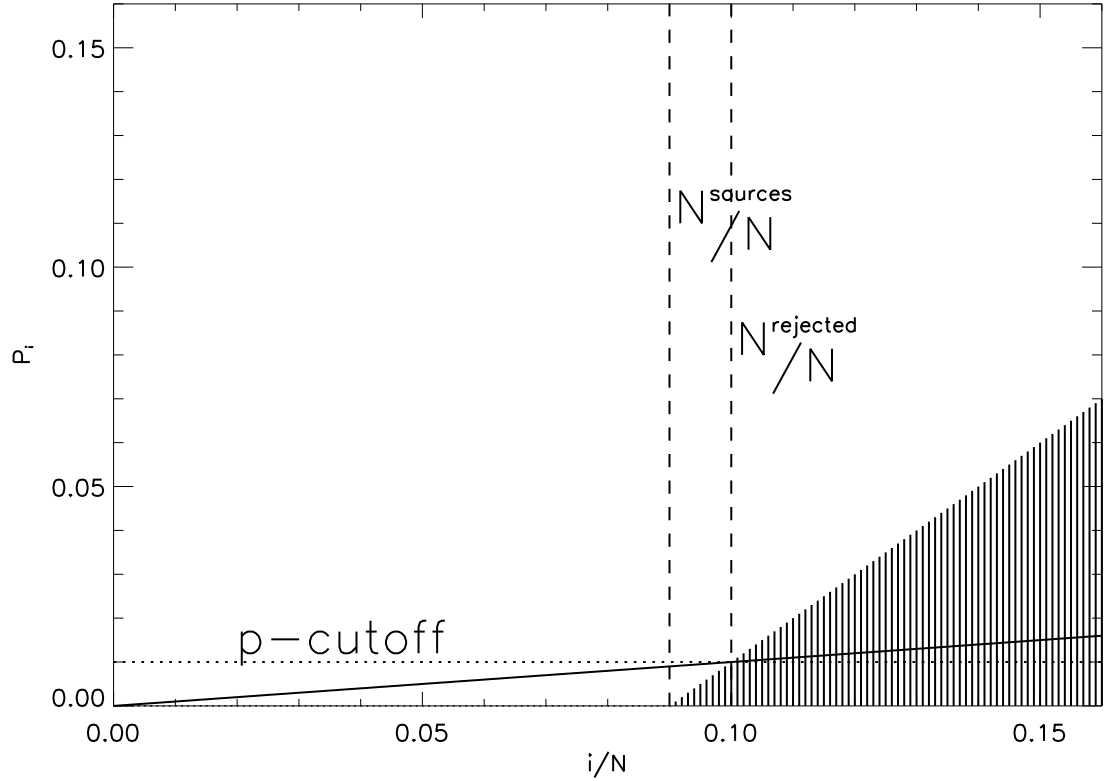}
\caption[]{\label{fig::app_pval} The ordered probabilities. The line has slope $\alpha= 0.10$. Notice where
the first crossing from the right occurs. Anything to the left of this is considered
rejected as a source pixel. Anything to the right is maintained as background. The
true source pixels have p-values set to zero.
The pixels between $N^{source}/N \le i/n \le N^{reject}/N$ are false
discoveries.  The number of false discoveries over the total number of rejections
is $\alpha$.
}
\end{figure}

\begin{figure}[p]
\epsfbox{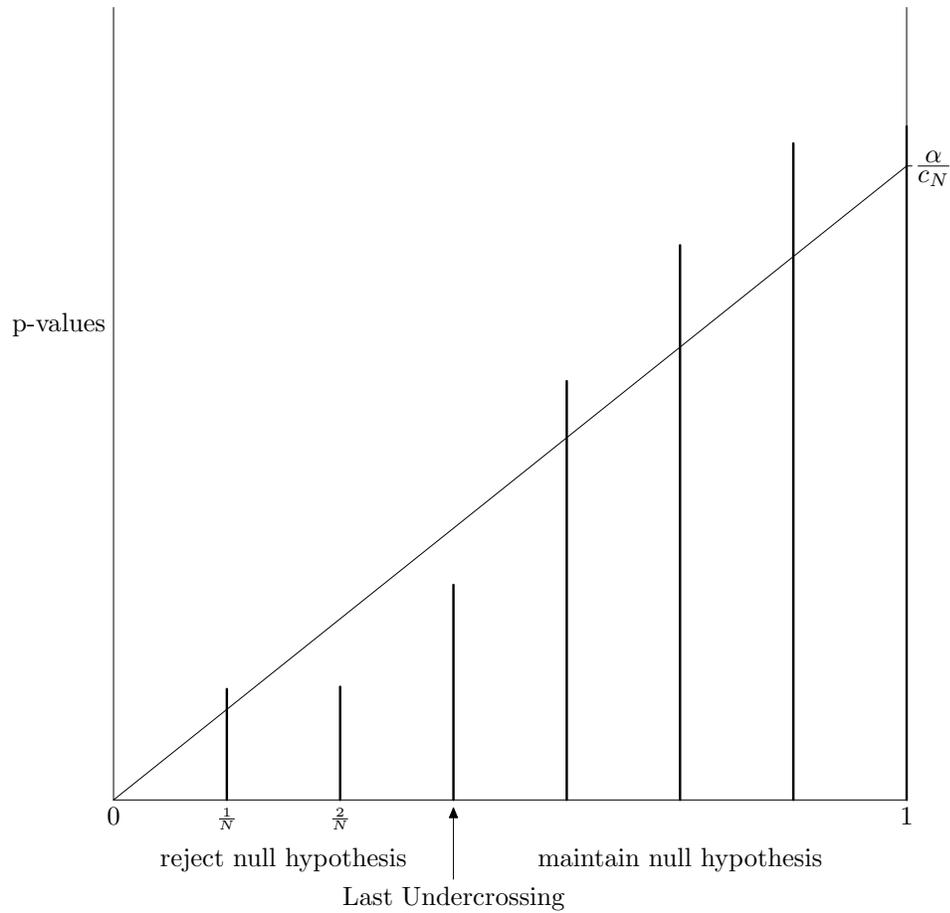}
\caption[]{\label{fig::fdr} 
A pictorial example of the FDR procedure applied to multiple hypothesis
tests. The vertical lines give the sorted p-values,
with the $j$th smallest p-value at horizontal position $j/N$.
The p-value at the last undercrossing (from the left) of the p-values below the line
of slope $\alpha/c_N$ determines which null hypotheses are rejected.
The null hypotheses for tests with p-values at or below this are rejected;
the null hypotheses for tests with p-values above this are maintained.}
\end{figure}

\end{document}